\documentstyle[12pt]{article}

\begin{document}

\begin{center}

\thispagestyle{empty}

\null
\vskip-1truecm

\hfill
IC/98/132

\vskip1truecm United Nations Educational Scientific and Cultural
Organization\\
and\\
International Atomic Energy Agency\\
\medskip THE ABDUS SALAM INTERNATIONAL CENTRE FOR THEORETICAL PHYSICS\\
\vskip1.5truecm
{\bf YANG-BAXTER EQUATION ON TWO-DIMENSIONAL LATTICE\\
AND SOME INFINITE DIMENSIONAL ALGEBRAS}
\vskip1.2truecm
M. Daoud\\

{\it
Laboratoire de Physique Th\'eorique (LPT), Facult\'e des Sciences,\\
Universit\'e Mohammed V, Av. Ibn Batouta, Rabat, Morocco,\\[1mm]
Facult\'e des Sciences, Universit\'e Ibram Zohr, Agadir, Morocco\\
and\\
The Abdus Salam International Centre for Theoretical Physics, Trieste,
Italy,
}\\[5mm]
J. Douari
\footnote{Junior Associate of the Abdus Salam ICTP.\\
\null\hspace{0.5cm} E-mail: douari@usa.net}\\

{\it Laboratoire de Physique Th\'eorique (LPT), Facult\'e des Sciences,\\
Universit\'e Mohammed V, Av. Ibn Batouta, Rabat, Morocco}\\[5mm]
and\\[5mm]
Y. Hassouni
\footnote{Regular Associate of the Abdus Salam ICTP.\\
\null\hspace{0.5cm} E-mail: y\_hassou@fsr.ac.ma}\\

{\it
Laboratoire de Physique Th\'eorique (LPT), Facult\'e des Sciences,\\
Universit\'e Mohammed V, Av. Ibn Batouta, Rabat, Morocco\\
and\\
The Abdus Salam International Centre for Theoretical Physics, Trieste,
Italy.}\\
\end{center}
\vspace{1cm}

\centerline{\bf Abstract}
\bigskip

We show that the Yang-Baxter equation is equivalent to the associativity of the
algebra generated by non-commuting link operators. Starting from these link
operators we build out the (FFZ) algebras, the $s\ell_q (2)$ is derived by
considering a special combination of the generators of (FFZ) algebra.
\vfill

\begin{center}
 MIRAMARE -- TRIESTE\\
\medskip
September 1998
\end{center}
\vfill

\newpage

\baselineskip=18pt

\section{Introduction}

\hspace*{.3in}The Yang-Baxter equation (YBE) is an unifying basis of several studies in
two-dimensional integrable systems described by the quantum inverse scattering
method [1]. A particular solution of this equation has lead to the definition
of quantum groups [2]. The latter are mathematical objects which arose in the
solution of some models of statistical mechanics [3] and in the study of
factorized scattering of solitons and strings [4].

One of the earlier discoveries in particle physics was the realization of the
existence of two different types of particles: bosons and fermions. In the
algebraic context they are distinguished by the fact that the bosonic
(fermionic) operators generally satisfy simple commutation (anti-commutation)
relations. In the context of two-dimensional quantum field theory, it is
natural
to accept that one would encounter more exotic objects than just bosons and
fermions. In fact, anyons which are two-dimensional particles with arbitrary
statistics, interpolate between bosons and fermions (for a review see [5]-[7]
and references therein).
In the last few years they have attracted a spectacular interest, especially in
the interpretation of certain condensed matter phenomena, most notably the
fractional quantum Hall effect [8] and high $T_c$-superconductivity [9].

Quantum groups present themselves as natural mathematical objects allowing
the description of the fractional statistics. Indeed, it has been proved in
several works that the connection between quantum deformations and intermediate
statistics holds [10-19].

The aim of this paper consists in obtaining the YBE by introducing some
non-commuting operators denoted by $L_\ell$. These ones link between different
sites of a given arbitrary two-dimensional lattice, we show that the
associativity of the algebra generated by $L_\ell$'s is equivalent to the YBE.
Starting from these link operators, we consider the anyonic algebra which
coincides exactly with the one obtained in the work [10]. We realize in a pure
mathematical context the Fairke-Fletcher-Zachos (FFZ) algebra, that can be
seen as a
quantum $W_\infty$ algebra [20]. We also derive the $s\ell_q (2)$ from the
(FFZ) algebra.

The article is organized as follows. In the second section we obtain the YBE by
requiring the non-commutativity of the link operators introduced on an
arbitrary two-dimensional lattice. In the third section we establish a
correspondence between the phase breaking the commutativity of the link
operators and the angle function on the two-dimensional lattice. This
construction leads to the anyonic algebra.

The fourth section is devoted to the introduction of the translation operators
on a lattice, we show that they are nothing but the generators of the (FFZ)
algebra. We derive the $s\ell_q (2)$ algebra. The final section consists on
giving some concluding remarks.

\section{The Yang-Baxter Equation (on two-dimensional lattice)}

\hspace*{.3in}This section is devoted to obtaining the Yang-Baxter equation starting from the
definition of some operator denoted by $L_\ell$. These ones allow the transition
from one site to an arbitrary other one on a given two-dimensional lattice. We
begin with the definition of the above operator and we show that the
Yang-Baxter equation is nothing but an equation which is equivalent to the
associativity of the algebra generated by the non-commuting elements $L_\ell$.

We denote by $\Omega$ a two-dimensional lattice and we define the link
operators $L_\ell$ as follows:

\paragraph{Definition 1}
$$\phi^\ell_i(n)\equiv L_\ell\phi_i(n)\eqno(1)$$
where $\phi_i(n)/i=1,\dots,d$ is a $d$-dimensional vector function on $\Omega$,
$n$ is defined to be a couple of two integer $n\equiv (n_1,n_2)$ ($n_1$ and
$n_2$ are respectively the horizontal and longitudinal coordinates of a given
site $n$ of $\Omega$).

In the equation (1) $\ell=\pm 1,\pm 2$ are the four possible orientations
on $\Omega$ (Fig.3), so the index ``$\ell$'' indicates $a$ moving along the
direction
$\ell$ on $\Omega$;

\paragraph{Definition 2}
$$\phi^1_i(n)\equiv \phi_i(n_1+1, n_2),\quad \phi^2_i(n)\equiv
\phi_i(n_1,n_2+1)$$
$$\phi^{-1}_i(n) \equiv \phi_i(n_1-1,n_2),\quad \phi^{-2}_i(n)\equiv
\phi_i(n_1,n_2-1)\eqno(2)$$
We note that in these definitions, the element $L_\ell$ is regarded as an
operator
linking two neighbouring lattice sites, so we call it the link operator. In the
simple case where these operators commute, one can write:
$$L^{(\ell')}_\ell\cdot L_{\ell'}=L^{(\ell)}_{\ell'}\cdot L_\ell\eqno(3)$$
we point out that in this relation, we adopt the notation $L^{(\ell')}_\ell$
for which we suppose that the link operator $L^{(\ell')}_\ell$ act on a given
vector function $\phi_i (n)$ in the direction $\ell$ by keeping invariant the direction described
by $\ell'$.

The product ``$\cdot$'' in Eq.(3) is simply the composition of the operators
$L_\ell$'s, this composition occurs from the definition (1). Now, by
introducing a
matrix denoted by $R_{\ell\ell'}$, we break the commutativity of the product (Eq.(3)) as
follows:
$$R_{\ell\ell'}:V\otimes V\to V\otimes V$$
and
$$(R_{\ell\ell'})^{ij}_{mn} \left( L^{(\ell')}_\ell\right)^m_{m'}\cdot
\left(L_{\ell'}\right)^n_{n'}=
\left(L^{(\ell)}_{\ell'}\right)^j_{n'}\cdot (L_\ell)^i_{m'}\eqno(4)$$
where $V$ is a $d$-dimensional vector space in which lives the multi-component
functions ``$\phi_i$''. The indices $i$, $j$, $m'$ and $n'$ take the values
$1,2,\dots,d$.

The equation (4) can be rewritten in a compact form as:
$$(R_{\ell\ell'})_{12}\left((L^{(\ell')}_\ell)_1\cdot
(L_{\ell'})_2\right)=(L^{(\ell)}_{\ell'})_2\cdot (L_\ell)_1\eqno(5)$$
owing to  this equality, one can prove, by a direct computation that the
relation:
$$(R_{\ell\ell'})_{12}(R_{\ell'\ell})_{21}=1\!1\ \otimes 1\!1\ \
.\eqno(6)$$
where $1\!1$ is the unit $d\times d$ matrix acting on $V$.

Starting from the above tools, we derive the following result:

\paragraph{Proposition 1}
By requiring the product in Eq.(3) to be associative, we obtain the well known
Yang-Baxter equation on $(R_{\ell\ell'})$:
$$(R_{\ell\ell'})_{12}(R^{(\ell')}_{\ell\ell''})_{13}(R_{\ell'\ell''})_{23}=(R^{
(\ell)}_{\ell'\ell''})_{23}
(R_{\ell\ell''})_{13}(R^{(\ell'')}_{\ell\ell'})_{12}\eqno(7)$$

 We recall that in the literature this equation is seen as
a representation of the braid group. The latter plays for the intermediate
quantum statistics the same role played by the permutation group for bosonic
and fermionic statistics. In mathematical sense, it suffices to multiply the
matrix $R$ in (7) by a permutation one $P$ as:
$$B=P\cdot R$$
and the YBE becomes:
$$B_{12} B_{23}B_{12}=B_{23}B_{12}B_{23}\eqno(8)$$
We notice that in the classical limit $R=1\!1 \ \otimes 1\!1$,
this equation becomes trivial.

The equality (8), known as the Braid relation appears in the study of
intermediate statistics (especially the anyonic ones). So, we construct, using
the above mathematical tools, an algebra, interpolating between the bosonic and
fermionic algebras. This matter constitutes the purpose of the next section.

\section{Anyonic Algebra}

\hspace*{.3in}Firstly, we point out that the approach leading to the anyon algebra is of
purely mathematical aspect. Moreover, it is treated in a way which is
different from those usually used in the literature.

We start by considering an elementary plaquette $P_n$ (Fig. 1). The operators $L_\ell$ along the
$P_n$ are written as:

\paragraph{Definition 3}

$$L_1\equiv e^{i A_1((n_1,n_2),(n_1+1,n_2))}$$
$$L_2\equiv e^{iA_2((n_1+1,n_2),(n_1+1,n_2+1))}\eqno(9)$$
$$L_{-1}\equiv e^{iA_1((n_1+1,n_2+1),(n_1,n_2+1))}$$
$$L_{-2}\equiv e^{iA_2((n_1,n_1+1),(n_1,n_2))}.$$
The functions $A_{\bar{l}(n,p)}$ are subject to the following
constraint:$$\partial_\ell f_{\overline\ell}
(n)=A_{\overline\ell}(n,n+\ell),\quad n\in\Omega\eqno(10)$$
In the relation (10), the derivative on $\Omega$ is defined by:
$$\partial_\ell f_{\bar\ell}(n)=f(n+\ell)-f(n)\eqno(11)$$
$$\partial_{-\ell}=-\partial_\ell$$

$\ell$ means the direction on the lattice and $\bar\ell$ is its absolute value.

The motivation for the choice of these link operators will be made clear
when we
give the construction of the anyonic algebra.

The phases in (9) may be rewritten as:
$$\begin{array}{rcl}
A_1((n_1,n_2),(n_1+1,n_2)) & = & f_1(n_1+1,n_2)-f(n_1,n_2)\\[3mm]
& = & \partial_1 f_1(n_1,n_2)\\[3mm]
&\equiv & f^1 (n_1,n_2)\\
A_2((n_1+1,n_2),(n_1+1,n_2+1)) & = & f_2(n_1+1,n_2+1)-f_2(n_1+1,n_2)\\[3mm]
& = & \partial_2 f_2(n_1,n_2)\\[3mm]
&\equiv& f^2(n_1+1,n_2)\\[3mm]
A_1((n_1+1,n_2+1)), (n_1,n_2+1)) & = & f_1(n_1,n_2+1)-f_1(n_{1}+1,n_2+1)\\[3mm]
& = & -\partial_1f_1(n_1,n_2)\\[3mm]
&\equiv & -f^1(n_1,n_2+1)\\[3mm]
A_2((n_1,n_2+1), (n_1,n_2)) & = & f_2(n_1,n_2)-f_2(n_1,n_2+1)\\[3mm]
& = & -\partial_2 f_1(n_1,n_2)\\[3mm]
& = & -f^2(n_1,n_2)
\end{array}
\eqno(13)$$

In this construction the $f^\alpha(n)$ can be interpreted as the angles between the sites $n_\alpha$ and $n_{\alpha}+1$, $\alpha=1,2$.
These angles are given via a point $n^*\in\Omega^*$ (we denote by
$\Omega^*$ the dual lattice of
$\Omega$ and $n^*\in\Omega^*$, $n^*\equiv(n_1+{a\over 2}, n_2+{a\over 2})$,
$a$ is the lattice
spacing between two neighbourhood sites).

It is natural, owing to Eqs.(13) to see that the sum of the four phases
over a plaquette $P_n$
in $\Omega$ is given by the relation:
$$f^2(n_1+1,n_2)-f^2(n_1,n_2)-f^1(n_1,n_2+1)+f^1(n_1,n_2)=2\pi\eqno(14)$$
So this expression can be rewritten for one tower as:
$$\oint_{P_n} f(n)=2\pi\eqno(15)$$
And thus for several turns one obtains:
$$\oint_{P_n} f(n)=2\pi \kappa\eqno(16)$$
where $\kappa$ is the winding number of the closed loop $P_n$. Thus in
general, for any curve
$\Gamma_n$ on the lattice we can define the function $\Theta(n)$ as:
$$\Theta_{\Gamma_n}(n)\equiv \int_{\Gamma_n} f(n),\quad n\in\Omega\eqno(17)$$
Such that $
\Gamma_n$ is the curve from a point $p$ at the infinity of the $x$-axis to
the point $n$ on the
lattice $\Omega$ (Fig.2). We call $\Theta_{\Gamma_n}(n)$ the lattice angle
function under which
the point $n$ may be regarded by another site $m$ on the lattice. The
relevance of angle function
appearing in the intermediate statistics has been introduced firstly in the
work [5]. According
to this work, the angle function is measured from another point
$m^*\in\Omega^*$ instead of
$n^*\in\Omega^*$ (Fig.2). We take therefore $\Theta_{\Gamma_n(n,m)}$ and
$f(n,m)$ instead of
$\Theta_{\Gamma_n(n)}$ and $f(n)$.

In the same way and accordingly of the above definitions one defines.

\paragraph{Definition 4}
$$\Theta_{\Gamma_n}(n,m)-\Theta_{\Gamma'_n}(n,m)\equiv
\oint_{\Gamma_n\Gamma'^{-1}_n} f(n,m)=2\pi
\kappa\eqno(18)$$
where $\kappa$ is the winding number of the loop $\Gamma_n\Gamma'^{-1}_n$
around the dual point
$m^*$ (Fig.2).

Basing on the fact, there are two kinds of the angle function, when one
considers the origin
point ``$p$'' at infinity) of the positive $x$-axes or at the infinity of
the negative $x$-axes.
One can prove that these functions are subject to the following conditions:
$$\Theta_{\pm\Gamma_n}(n,m)-\Theta_{\pm\Gamma_m}(m,n)=\left\{
\begin{array}{l}
\pm\prod{\rm sgn}(n_2-m_2),\ n_2\not= m_2\\[3mm]
\pm\prod {\rm sgn} (n_1-m_1),\ n_2=m_2\end{array}\right.$$
$$\Theta_{-\Gamma_n}(n,m)-\Theta_{\pm\Gamma_n}(n,m)=\left\{
\begin{array}{l}
-\prod{\rm sgn}(n_2-m_2),\ n_2\not= m_2\\[3mm]
-\prod {\rm sgn} (n_1-m_1),\ n_2=m_2\end{array}\right.\eqno(19)$$
with ``$+\Gamma_n$'' the curve following the anti-clockwise sense and
``$-\Gamma_n$'' is the
clockwise one.

Now, we are in a position to construct the anyonic operators. They are given
starting from the
introducing of fermionic field on a two-dimensional lattice. Let us give
the two-component
fermionic spinor field by:
$$\psi(n)=\left(\begin{array}{c} \psi_1(n)\\[3mm]
\psi_2(n)\end{array}\right)\eqno(20)$$
The quantized components in relation (20) obey to the following equation:
$$\{\psi_\alpha(n),\ \psi_\beta(m)\}=0$$
$$\{\psi_\alpha(n),\
\psi^\dagger_\beta(m)\}=\delta_{nm}\delta_{\alpha\beta}\eqno(21)$$
$$\{\psi^\dagger_\alpha(n),\ \psi^\dagger_\beta(m)\}=0$$
where $\alpha,\beta=1,2$ and $\{\varphi,\eta\}\equiv \varphi\eta+\eta\varphi$.

Let us recall also that the Fock vacuum state $\vert 0\rangle$ is defined by:
$$\psi_\alpha(n)\vert 0\rangle = 0\eqno(22)$$
starting from this vacuum, one can generate all the other states describing
the fermions on
$\Omega$.

Returning to our purpose; the construction of the anyonic algebra. To start
let us at first
introduce the elements:
$$\Delta_\alpha(n_\pm)=\sum_m\psi^\dagger_\alpha(m) \Theta_{\pm\Gamma_n}(n,m)\
\psi_\alpha(m)\eqno(23)$$
By straightforward calculation, one obtains
$$[\Delta_\alpha(n_\pm),\ \psi_\beta(m)]=-\delta_{\alpha\beta}\Theta_{\pm
\Gamma_n}(n,m)\
\psi_\alpha(m)$$
$$[\Delta_\alpha(n_\pm),\ \psi^\dagger_\beta(m)]=\delta_{\alpha\beta}\Theta_{\pm
\Gamma_n}(n,m)\psi^\dagger_\alpha(m)\eqno(24)$$
$$[\Delta_\alpha(n_\pm),\ \Delta_\beta(m_\pm)]=0$$

Now we can give the expression of the anyonic operators. Indeed one can
prove the following
proposition:

\paragraph{Proposition 2}
By defining the operators $\varphi_\alpha(n_\pm)$ and
$\varphi^\dagger_\alpha(n_\pm)$ as:
$$\varphi_\alpha(n_\pm)=e^{i\nu\Delta_\alpha(n_\pm)} \psi_\alpha(n)$$
$$\varphi^\dagger_\alpha(n_\pm)
=\psi^\dagger_\alpha(n)e^{-i\nu\Delta_\alpha(n_\pm)}\eqno(25)$$

We obtain the algebra described by the following algebraic relations:
$$\{\varphi_\alpha(n_\pm),\ \varphi^\dagger_\alpha(n_\pm)\}=1$$
$$\{\varphi_\alpha(n_\pm),\ \varphi_\alpha(m_\mp)\}_{\Lambda^\mp}=0\qquad
n>m$$
$$\{\varphi_\alpha(n_\pm),\ \varphi^\dagger_\alpha(m_\pm)\}_{\Lambda^\mp}=0\qquad
n>m$$
$$\{\varphi^\dagger_\alpha(n_\pm),\ \varphi_\alpha(m_\pm)\}_{\Lambda^\mp}=0\qquad
n>m\eqno(26)$$
$$\{\varphi^+_\alpha(n_\pm),\
\varphi^\dagger_\alpha(m_\mp)\}_{\Lambda^\mp}=0\qquad  n>m$$
$$\{\varphi_\alpha(n_\pm),\ \varphi_\beta(m_\pm)\}=0\qquad  \alpha\not=\beta$$
$$\{\varphi^\dagger_\alpha(n_\pm), \ \varphi_\beta(m_\pm)\}=0\qquad
\alpha\not=\beta$$
$$\{\varphi^\dagger_\alpha(n_\pm),\ \varphi^\dagger_\beta(m_\pm)\}=0\qquad
\alpha\not=\beta$$
where $\Lambda^\pm =e^{\pm i\nu\pi}$ and $n>m\Leftrightarrow
\left\{\begin {array}{l}
n_+>m_+\Leftrightarrow\left\{\begin{array}{l} n_2>m_2\\[2mm]
n_1>m_1, n_2-n_1\end{array}\right.\\[3mm]
n_-<m_-\Leftrightarrow
\left\{\begin{array}{l}
n_2<m_2\\[2mm]
n_1<m_1\\[2mm]
n_2=m_2\end{array}\right.\end{array}\right.$
and $\{x,y\}_\Lambda = xy+\Lambda yx$.

We have also
$$[\varphi_\alpha(n_\pm)]^2=[\varphi^\dagger_\alpha(n_\pm)]^2=0\eqno(27)$$

This constraint seems to describe an important property of anyonic systems.
Indeed it appears
when one discusses the statistics corresponding to anyons, the relation
Eq.(27) is viewed as a
hard core condition; at the same point of a two-dimensional lattice, can not
live more than one
particle. For this reason, many of the authors in the literature consider
the anyons as a fermion
but defined on a given two-dimensional lattice. We add also to this remark
that, owing to the
above condition, anyons obey the Pauli exclusion principle.

The parameter $\nu$ in the equalities (25) is seen as a statistical
parameter. The obtained
algebra interpolates between the bosonic algebra $(\nu=1\ {\rm mod}\ 2$)
and the fermionic one
$(\nu=0\ {\rm mod}\ 2)$.

Consequently, we have realized the anyonic algebra starting from one
special definition of the
link operators on the two-dimensional lattice $\Omega$. We showed also the
correspondance between
them and the angle function discussed in [5]. In the latter, the authors
show that the Schwinger
realization of SU(2) which is bosonic or fermionic, can be generalized to
anyons of intermediate
statistics. In this case, the Schwinger construction does not lead to
ordinary group SU(2), but
rather to its $q$-analogue, the $U_q(2)$ $(q=\exp i \pi\nu$).

In order to investigate some other quantum symmetries appearing when the
studying of these exotic
statistics, we will show that it is possible to obtain the FFZ algebra from
which we derive the
quantum group $s\ell q(2)$.

\section{(FFZ) and $s\ell_q (2)$ algebras}

\hspace*{.3in}Starting from the above link operators (Eq.(9)), we define the generators
$T_n$ as follows:

\paragraph{Definition 5}
$$T_n=T_{(n_1,n_2)}=R_{ij}^{\bar n_1\bar n_2\over 2} L^{\bar n_i}_{\pm i}
L^{\bar n_j}_{\pm
j}\eqno(28)$$
where $n_1,n_2\in\bf{Z}\rm$, $i,j=1,2$ and $i\not= j$. $\bar n$ is the
absolute value of $n\in\bf{Z}\rm$, the indices are omitted in the notation of $T_{(n_1,n_2)}$.

As already seen (Fig.1) the operator $L_\ell$ allows the transition from
the site $X_i$ to the
site $X_{i+\ell}$ on $\Omega$. This is described in Fig.3,
$\ell=\pm1,\pm2$. The generators
$T_n$ are regarded as translations from the point O (the origin of
referential) to a point
$n\in\Omega$ (Fig.4).

At first we require that the product of these operators is given by formula:
$$T_nT_m=e^{i\alpha_2(n,m)}T_{n+m}\eqno(29)$$
The function $\alpha_2 (n,m)$ depending on two sites of $\Omega$ is introduced
to be anti-symmetric
and is defined as:
$$\alpha_2:\Omega\times\Omega\to\bf{R}\rm.$$ The motivation of this choice of the product between these two translation
operators is due to
the fact that the link operators do not commute and thus the composition of
two translations
must lead naturally to a translation.

Before giving the complete description of the (FFZ), we notice that in the
literature this
algebra has been poorly realized in a mathematical way. So, one of the main
results of this work
is to construct the FFZ algebra on the lattice $\Omega$. To do this we are
lead, due to a pure
mathematical reason, to divide $\Omega$ onto four subsets given by:
$$i)\quad D^{++};\ n\in D^{++}\Leftrightarrow (n_2>0,\ n_1>0)$$
$$ii)\quad D^{+-};\ n\in D^{+-}\Leftrightarrow (n_2>0, n_1<0)$$
$$iii)\quad D^{-+};\ n\in D^{-+}\Leftrightarrow (n_2<0, n_1>0)\eqno(30)$$
$$iv)\quad D^{--};\ n\in D^{--}\Leftrightarrow(n_2<0,\ n_1<0)\ .$$
and
$$\Omega=D^{++}\oplus D^{+-}\oplus D^{-+}\oplus D^{--}\ .$$
Now, by requiring that the translation operators $T_n$ do not commute, but
this non-commutativity
property is described by the introducing of some matrix $R$ as follows:
$$T_nT_m=R^{ab}_{nm} T_aT_b\eqno(30a)$$
we find that the matrix obey, owing to the assumption (Eq.(29)), the
following relation:

\paragraph{Proposition 3}
$$R^{ab}_{nm} =\delta^{a+b}_{n+m}
e^{-i(\alpha_2(a,b)-\alpha_2(m,n))}\eqno(30b)$$
one can check that this matrix satisfies the Yang-Baxter Equation, because
the product composing
the translation operators in the expression (Eq.(30b)) is associative.

At this stage we are able to give the expression of the generators of the
(FFZ) algebra on every
subset of the lattice $\Omega$.\\
\bigskip
\noindent{\bf Definition 6}\\ \bigskip i) for $D^{++}$: $$T^{++}_n\equiv R^{\bar n_2\bar n_1\over 2}_{2,1} L^{\bar n_2}_2 L^{\bar n_1}_1$$ $$S^{++}_n\equiv R^{\bar n_2\bar n_1\over 2}_{1,2} L^{\bar n_1}_1 L^{\bar
n_2}_2\eqno(31a)$$
ii) for $D^{+-}$:
$$T^{+-}_n\equiv R^{\bar n_2\bar n_1}_{1,2} L^{\bar n-1}_{-1} L^{\bar n-2}_2$$
$$S^{+-}_n\equiv R^{\bar n_2\bar n_1\over 2}_{2,1} L^{\bar n_1}_2 L^{\bar
n_2}_{-1}\eqno(31b)$$
iii) for $D^{-+}$:
$$T^{-+}_n\equiv  R^{\bar n_2\bar n_1\over 2}_{1,2} L^{\bar n_1}_1 L^{\bar
n_2}_{-2}$$
$$S^{-+}_n\equiv R^{\bar n_2\bar n_1\over 2}_{2,1} L^{\bar n_2}_{-2} L^{\bar
n_1}_1\eqno(31c)$$
iv) for $D^{--}$:
$$T^{--}_n\equiv R^{\bar n_2\bar n_1\over 2} L^{\bar n_2}_{-2} L^{\bar
n_1}_{-1}$$
$$S^{--}_n \equiv R^{\bar n_2\bar n_1\over 2}_{1,2} L^{\bar n_1}_{-1} L^{\bar
n_2}_{-2}\eqno(31d)$$

Following the assumption (Eq.(30)), we prove by a direct calculation that
the parameters
$R_{1,2}$ appearing in the relations (31) have the expression:
$$R_{1,2} =e^{i\alpha_{2}(n,n')\over \bar n'\times\bar n}$$

Consequently, the elements ($T$'s) are nothing but the generators of (FFZ)algebra on the lattice $\Omega$, we have then:
\paragraph{Proposition 4}

$${\rm for}\ D^{++};\quad [T^{++}_n,T^{++}_{n'}]=2i\sin
\alpha(n,n')T^{++}_{n+n'}$$
$${\rm for}\ D^{+-};\quad [T^{+-}_n,T^{+-}_{n'}]=2i\sin
\alpha(n,n')T^{+-}_{n+n'}$$
$${\rm for}\ D^{-+};\quad [T^{-+}_n,T^{-+}_{n'}]=2i\sin
\alpha(n,n')T^{-+}_{n'+n}$$
$${\rm for}\ D^{--};\quad [T^{--}_n,T^{++}_{n'}]=2i\sin
\alpha(n,n')T^{--}_{n+n'}\eqno(32)$$
and the same results for the elements $S$'s.

Using these particular realizations of (FFZ) algebra, we can derive the
$s\ell_q (2)$ starting
from the generators defined by:

\paragraph{Definition 7}
$$J_+\equiv {1\over (q-q^{-1})}(T^{++}_{(1,1)}-T^{+-}_{(-1,1)})$$
$$J_-\equiv {1\over (q-q^{-1})}(T^{--}_{(-1,-1)}-T^{-+}_{(1,-1)})\eqno(33)$$
$$q^{2J_3}\equiv T^{++}_{(2,0)}$$
$$q^{-2J_3}\equiv T^{--}_{(-2,0)}$$

Basing on these definitions, we obtain:
$$[J_+,J_-]=[2J_3]_q$$
$$q^{J_3}J_\pm q^{-J_3}=q^{\pm 1}J_\pm\eqno(34)$$
where $q$ is taken to equal $R_{1,2}=R_{2,1}$ in the above equations and
$$[x]_q={q^x-q^{-x}\over q-q^{-1}}\ .$$

We point out that surprisingly enough, we have constructed the $s\ell_q (2)$
algebra starting from
the introduction of the link operators on a lattice. Our
realization is different
than the one given in the work [5] where the authors obtain the same
quantum symmetry by using
the Schwinger construction.

\section{Concluding Remarks}

\hspace*{.3in}In some mathematical point of view, the notion of discretization of the
two-dimensional manifold
has been seen in the literature as the main and essential conception
allowing the connection
between the intermediate statistics and quantum algebras. In this context,
starting from the
definition of the link operators on a given two-dimensional lattice
$\Omega$, we have constructed
an algebra coinciding exactly with the anyonic algebra. We have realized,
in a mathematical way
the (FFZ) algebra build out from the introduced link operators on
$\Omega$. The $s\ell q(2)$ is
thus obtained in an original way.

\section*{Acknowledgments}

\hspace*{.3in}Two of the authors (M.D. and Y.H.) would like to thank Professors M.
Virasoro and G.
Ghirardi, the International Atomic Energy Agency and UNESCO for hospitality
at the
Abdus Salam International Centre for Theoretical Physics, Trieste. A
particular thanks goes to
Professor G. Thompson for his helpful and interesting discussions and comments.
This work was done within the framework of the Associateship Scheme of the
Abdus
Salam International Centre for Theoretical Physics, Trieste.

\newpage
\baselineskip=18pt

$$\begin{picture}(0,0)
\put(-30,0){\line(0,-1){90}}
\put(20,0){\line(0,-1){90}}
\put(-50,-20){\line(1,0){90}}
\put(-50,-70){\line(1,0){90}}
\put(-33,-23){$\bullet$}
\put(-33,-73){$\bullet$}
\put(17,-23){$\bullet$}
\put(17,-73){$\bullet$}
\put(-10,-23){$\leftarrow$}
\put(-10,-73){$\rightarrow$}
\put(17,-50){$\uparrow$}
\put(-33,-50){$\downarrow$}
\put(-15,-15){$L_{-1}$}
\put(-15,-80){$L_{1}$}
\put(-52,-45){$L_{-2}$}
\put(25,-45){$L_{2}$}
\put(-68,-80){$(n_1,n_2)$}
\put(-200,-110){{\bf Fig.1} -- Elementary plaquette $A_n$,
$n=(n_1,n_2)\in\Omega$, on which we
translate from $n$ to $n$.}
\end{picture}$$

$$\begin{picture}(0,0)
\put(-75,-130){\thinlines\line(1,0){145}}
\put(-75,-145){\thinlines\line(1,0){145}}
\put(-75,-160){\thinlines\line(1,0){145}}
\put(-75,-175){\thinlines\line(1,0){145}}
\put(-75,-190){\thinlines\line(1,0){145}}
\put(-75,-205){\thinlines\line(1,0){145}}
\put(-75,-220){\thinlines\line(1,0){145}}
\put(-70,-125){\thinlines\line(0,-1){100}}
\put(-55,-125){\thinlines\line(0,-1){100}}
\put(-40,-125){\thinlines\line(0,-1){100}}
\put(-25,-125){\thinlines\line(0,-1){100}}
\put(-10,-125){\thinlines\line(0,-1){100}}
\put(5,-125){\thinlines\line(0,-1){100}}
\put(20,-125){\thinlines\line(0,-1){100}}
\put(35,-125){\thinlines\line(0,-1){100}}
\put(50,-125){\thinlines\line(0,-1){100}}
\put(65,-125){\thinlines\line(0,-1){100}}
\put(-55,-145){\linethickness{0.5mm}\line(1,0){45}}
\put(-10,-145){\linethickness{0.5mm}\line(0,-1){15}}
\put(-10,-160){\linethickness{0.5mm}\line(1,0){15}}
\put(20,-160){\linethickness{0.5mm}\line(1,0){45}}
\put(65,-160){\linethickness{0.5mm}\line(0,-1){30}}
\put(35,-190){\linethickness{0.5mm}\line(1,0){30}}
\put(35,-190){\linethickness{0.5mm}\line(0,-1){30}}
\put(-55,-145){\linethickness{0.5mm}\line(0,-1){30}}
\put(-70,-175){\linethickness{0.5mm}\line(1,0){15}}
\put(-70,-175){\linethickness{0.5mm}\line(0,-1){30}}
\put(-70,-205){\linethickness{0.5mm}\line(1,0){30}}
\put(-40,-205){\linethickness{0.5mm}\line(0,-1){15}}
\put(-40,-220){\linethickness{0.5mm}\line(1,0){60}}
\put(-38,-147){$\leftarrow$}
\put(-8,-163){$\leftarrow$}
\put(52,-163){$\leftarrow$}
\put(37,-193){$\leftarrow$}
\put(-23,-223){$\leftarrow$}
\put(-53,-208){$\leftarrow$}
\put(-68,-178){$\rightarrow$}
\put(-57,-156){$\uparrow$}
\put(-12,-156){$\uparrow$}
\put(63,-170){$\uparrow$}
\put(-73,-200){$\uparrow$}
\put(33,-215){$\downarrow$}
\put(63,-183){$\uparrow$}
\put(37,-155){$\Gamma_n$}

\put(75,-189){$p$}
\put(61,-191){$\bullet$}
\put(-65,-140){$n$}
\put(-59,-148){$\bullet$}
\put(-38,-185){$m^*$}
\put(-22,-185){$\bullet$}
\put(-39,-198){$m^*$}
\put(-28,-191){$\bullet$}
\put(-10,-180){\oval(14,14)[tr]}
\put(-20,-176){$\leftarrow$}
\put(-14,-171){$\Theta_{\Gamma_n(n,m)}$}
\put(80,-120){\oval(14,14)[tr]}
\put(72,-116){$\leftarrow$}
\put(75,-111){$+$}
\put(-10,-240){$\Gamma'_n$}
\put(-200,-260){{\bf Fig.2} - $\Theta_{\Gamma_n(n,m)}$ is the angle under
which the point $n$
is seen by $m^*$ in the positive}
\put(-200,-280){direction (+) and $\Theta_{\Gamma'_n(n,m)}$ in the
opposite direction$(-)$}
\end{picture}$$

\newpage

$$\begin{picture}(0,0)

\put(-95,-145){\thinlines\line(1,0){200}}

\put(-95,-115){\thinlines\line(1,0){200}}

\put(-95,-85){\thinlines\line(1,0){200}}

\put(-95,-55){\thinlines\line(1,0){200}}

\put(-95,-25){\thinlines\line(1,0){200}}

\put(-70,-5){\thinlines\line(0,-1){160}}

\put(-40,-5){\thinlines\line(0,-1){160}}

\put(-10,-5){\thinlines\line(0,-1){160}}

\put(20,-5){\thinlines\line(0,-1){160}}

\put(50,-5){\thinlines\line(0,-1){160}}

\put(80,-5){\thinlines\line(0,-1){160}}

\put(-120,-185){\vector(0,1){170}}
\put(-120,-185){\vector(1,0){220}}

\put(-135,-200){$O$}
\put(-135,-30){$Y$}
\put(90,-200){$X$}
\put(-110,-220){{\bf Fig.3} - Lattice $\Omega$ and its reference XOY}

\put(-5,-50){$X_{i+2}$}
\put(-13,-58){$\bullet$}
\put(-13,-88){$\bullet$}
\put(17,-88){$\bullet$}
\put(-43,-88){$\bullet$}
\put(-13,-118){$\bullet$}

\put(-11,-86){\vector(0,1){20}}
\put(-11,-85){\vector(1,0){20}}
\put(-11,-85){\vector(-1,0){20}}
\put(-11,-85){\vector(0,-1){20}}
\put(-26,-75){$L_2$}
\put(6,-77){$L_1$}
\put(-64,-80){$X_{i-1}$}
\put(26,-80){$X_{i+1}$}
\put(120,-50){$(\Omega)$}
\put(-33,-98){$L_{-1}$}
\put(-6,-109){$L_{-2}$}
\put(-6,-130){$X_{i-2}$}

\end{picture}$$

$$\begin{picture}(0,0)
\unitlength=1cm
\put(-6,-9){\line(1,0){12}}
\put(-6,-18){\line(1,0){12}}
\put(-6,-9){\line(0,-1){9}}
\put(6,-9){\line(0,-1){9}}
\put(0,-8){\line(0,-1){11}}
\put(-7,-13.5){\line(1,0){14}}

\put(-0.5,-12.5){\line(-1,0){1.8}}
\put(-0.5,-12.2){\line(-1,0){1.8}}
\put(-0.5,-11.9){\line(-1,0){1.8}}
\put(-0.5,-11.6){\line(-1,0){1.8}}
\put(-0.8,-12.8){\line(0,1){1.5}}
\put(-1.1,-12.8){\line(0,1){1.5}}
\put(-1.4,-12.8){\line(0,1){1.5}}
\put(-1.7,-12.8){\line(0,1){1.5}}
\put(-2,-12.8){\line(0,1){1.5}}

\put(2.3,-12.5){\line(-1,0){1.8}}
\put(2.3,-12.2){\line(-1,0){1.8}}
\put(2.3,-11.9){\line(-1,0){1.8}}
\put(2.3,-11.6){\line(-1,0){1.8}}
\put(2,-12.8){\line(0,1){1.5}}
\put(1.7,-12.8){\line(0,1){1.5}}
\put(1.4,-12.8){\line(0,1){1.5}}
\put(1.1,-12.8){\line(0,1){1.5}}
\put(0.8,-12.8){\line(0,1){1.5}}

\put(-0.5,-14.6){\line(-1,0){1.8}}
\put(-0.5,-14.9){\line(-1,0){1.8}}
\put(-0.5,-15.2){\line(-1,0){1.8}}
\put(-0.5,-15.5){\line(-1,0){1.8}}
\put(-0.8,-15.8){\line(0,1){1.5}}
\put(-1.1,-15.8){\line(0,1){1.5}}
\put(-1.4,-15.8){\line(0,1){1.5}}
\put(-1.7,-15.8){\line(0,1){1.5}}
\put(-2,-15.8){\line(0,1){1.5}}

\put(2.3,-14.6){\line(-1,0){1.8}}
\put(2.3,-14.9){\line(-1,0){1.8}}
\put(2.3,-15.2){\line(-1,0){1.8}}
\put(2.3,-15.5){\line(-1,0){1.8}}
\put(0.8,-15.8){\line(0,1){1.5}}
\put(1.1,-15.8){\line(0,1){1.5}}
\put(1.4,-15.8){\line(0,1){1.5}}
\put(1.7,-15.8){\line(0,1){1.5}}
\put(2,-15.8){\line(0,1){1.5}}

\put(2,-14.5){\oval(1.3,1.3)[tr]}
\put(2.65,-14.5){\vector(0,-1){0.2}}

\put(-2,-14.5){\oval(1.3,1.3)[tl]}
\put(-2.66,-14.5){\vector(0,-1){0.2}}

\put(-1,-15.6){\oval(1.3,1.3)[br]}
\put(-1,-16.26){\vector(-1,0){0.2}}

\put(1,-15.6){\oval(1.3,1.3)[bl]}
\put(1,-16.26){\vector(1,0){0.2}}

\put(1,-11.5){\oval(1.3,1.3)[tl]}
\put(1,-10.85){\vector(1,0){0.2}}

\put(2,-12.5){\oval(1.3,1.3)[br]}
\put(2.65,-12.5){\vector(0,1){0.2}}

\put(-0.9,-11.6){\oval(1.3,1.3)[tr]}
\put(-0.9,-10.95){\vector(-1,0){0.2}}

\put(-2,-12.5){\oval(1.3,1.3)[bl]}
\put(-2.66,-12.5){\vector(0,1){0.2}}

\put(-0.6,-8.5){$Y$}
\put(-5.5,-18.7){$(D^{--})$}
\put(4.5,-18.7){$D^{-+}$}

\put(-6.1,-9.1){$\bullet$}
\put(5.9,-9.1){$\bullet$}
\put(-6.1,-18.1){$\bullet$}
\put(5.9,-18.1){$\bullet$}

\put(-6.5,-9){$m$}
\put(6.2,-9){$n$}

\put(6.2,-9){\oval(1.3,1.3)[tr]}
\put(6.2,-8.35){\vector(-1,0){0.2}}
\put(7,-8.5){$+$}

\put(-6.5,-18.2){$q$}
\put(6.2,-18.2){$p$}

\put(6.2,-13.5){\vector(1,0){1}}
\put(6.7,-14){$X$}

\put(6.2,-10.5){$D^{++}$}
\put(-7.4,-10.5){$(D^{+-})$}

\put(-1,-10.5){$T^{+-}_m$}
\put(0.5,-10.5){$S^{++}_n$}

\put(-3.5,-13){$S^{+-}_m$}
\put(2.8,-13){$T^{++}_n$}

\put(-3.5,-14.2){$T^{--}_q$}
\put(-1,-16.8){$S^{--}_q$}

\put(3,-14.2){$S^{-+}_p$}
\put(0.3,-16.8){$T^{-+}_p$}

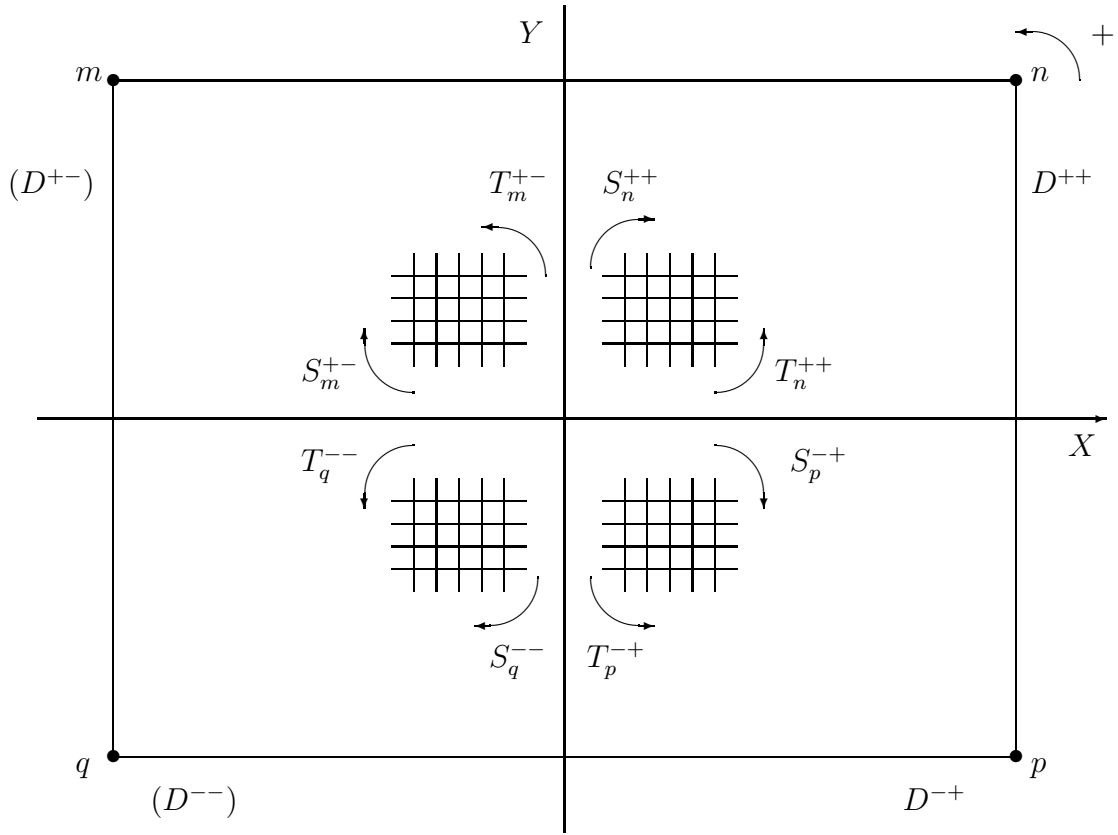
\put(-5.5,-20){{\bf Fig.4} - Lattice $\Omega$ is divided in four parts:
$D^{++}$, $D^{+-}$,
$D^{-+}$ and $D^{--}$}

\end{picture}$$

\end{document}